\documentclass[useAMS,usenatbib,usegraphicx,referee]{mn2e}

\usepackage{aas_macros}
\usepackage{longtable}

\title[How Good is the Cloud Cover Forecast?]{Ultimate Meteorological Question from Observational Astronomers: How Good is the Cloud Cover Forecast?}
\author[Q.-Z.~Ye \& S.-S.~Chen]
  {Q.-Z.~Ye\thanks{E-mail: qye22@uwo.ca},$^{1,2}$ and S.-S.~Chen$^{1,3}$ \\
  $^{1}$Department of Atmospheric Sciences, School of Environmental Science and
Engineering, Sun Yat-sen University, Guangzhou, China \\
  $^{2}$Department of Physics and Astronomy, The University of Western Ontario,
London, Ontario, N6A 3K7 Canada \\
  $^{3}$Department of Atmospheric and Oceanic Sciences, McGill University,
Montreal, Quebec, H3A 0B9 Canada}
\begin{document}

\date{Accepted 1970 January 1. Received 1970 January 1; in original form 1970
January 1}

\pagerange{\pageref{firstpage}--\pageref{lastpage}} \pubyear{1970}

\maketitle

\label{firstpage}

\begin{abstract}
To evaluate the capability of numerical cloud forecast as a meteorological reference for astronomical observation, we compare the cloud forecast from NCEP Global Forecast System (GFS) model for total, layer and convective cloud with normalized satellite observation from the International Satellite Cloud Climatology Project (ISCCP), for the period of July 2005 to June 2008. In general, the model forecast is consistent with the ISCCP observation. For total cloud cover, our result shows the goodness of the GFS model forecast with a mean error within $\pm15\%$ in most areas. The global mean probability of $<30\%$ forecast error (polar regions excluded) declines from 73\% to 58\% throughout the $180h$ forecast period, and is more skilled than ISCCP-based climatology forecast up to $\tau\sim120h$. The comparison on layer clouds reveals a distinct negative regional tendency for low cloud forecast and a questionable positive global tendency for high cloud forecast. Fractional and binary comparisons are performed on convective cloud forecast and revealed the GFS model can only identify less than half of such cloud. In short, our result suggests that the GFS model can provide satisfactory worldwide total cloud forecast up to a week ahead for observation scheduling purpose, but layer and convective cloud forecast is less reliable than the total cloud forecast.
\end{abstract}

\begin{keywords}
atmospheric effects, site testing.
\end{keywords}

\section{Motivation}

The most important prerequisite of most successful astronomical observation, mainly optical observation, is no doubt a cloudless sky \citep[see the story of Guillaume Le Gentil in the 1761/69 Transit of Venus, cf.][for a rather unlucky example]{saw51}. However, cloud cover forecast had been difficult for a long time as limited by theoretical understanding of mesoscale circulation (i.e. modeling) and computation ability. Numerical simultaneous cloud cover forecast for astronomical observation had not been practically useful until the very end of 20th century.

In the meteorological context, cloud plays a key role in radiation balance of the Earth, and is widely accepted to be the main source of uncertainty of global weather predictions \citep[cf.][]{stu99,ste05}. It gathered much attention from atmospheric sciences community, both from modeling group and from observational group. The attempt to ``parameterize'' cloud activity started in the 1980s \citep[cf.][and references therein]{sun89}. Several cloud schemes have been proposed since then, followed by ground- and space-based evaluation in aim for their refinement \citep[e.g.][]{hin99,luo05,yan06}.

Although many amateur and professional observatories have been taken advantage of open access to the outputs of major numerical weather models for a decade, the reliability of such forecast is poorly understood. The meteorology community is mostly interested in particular mesoscale events and/or particular regions, and pays relatively little attention to the model performance over broader environment; while there are only two studies from the astronomy community that have investigated this topic: \citet{era01}, which suggested that only 15-25\% of cloudy nights at the European Southern Observatory sites could be identified by the European Centre for Medium-Range Weather Forecasting (ECMWF) model; and our earlier study \citep{ye11}, which suggested a high detection rate accompanied with a moderate false alarm rate from the NCEP Global Forecast System (GFS) model, based on the cloud observations from several astronomical observatories. Even so, as the two studies are both limited at spatial densities and 
scales, investigations of 
numerical cloud forecasts over global scale are still lacking.

This study is therefore carried out in aim to assess the cloud forecast ability of a global numerical model to provide insight on its reliability as a reference for astronomical observation. In order to do this, we need: i) output from global numerical model as forecast data, and ii) observational data with appreciable temporal and spatial coverage. The selection and reduction of such data will be discussed in the following section.

\section{Data Selection and Preparation}

\subsection{Modeling Data}

The ECMWF and GFS models mentioned in \S1 are among the major global numerical weather models used for daily weather forecast. In this study, we follow our earlier study and use the GFS model as well. Firstly, we retrieve the GFS data in grid 003 ($1^{\circ}\times1^{\circ}$ grid) for the period of July 2005 to June 2008 from the National Operational Model Archive \& Distribution System \citep[NOMADS; see][]{rut06}. The data is in 3-hourly interval for forecast lead time ($\tau$) up to 180h and is produced four times per day at 00Z, 06Z, 12Z and 18Z (we refer each of them as one ``initialization'' hereafter). The data is then weight-averaged into the spatial resolution of $2.5^{\circ}$ to match the definition of the observational data that to be described in \S2.2.

The forecast cloud fraction in decimals, $C$, is computed using the Xu-Randall cloud scheme \citep{xu96}:

\begin{displaymath}
C=RH^{k_1} \times \left\{ 1 - \exp \left[ -\frac{k_2 q_l}{((1-RH)q_s)^{k_3}} \right] \right\}
\end{displaymath}

\noindent
where $RH$ is the environmental relative humidity, $q_l$ is the liquid water mixing ratio, $q_s$ is the saturation specific humidity, and $k_1=0.25$, $k_2=100$, and $k_3=0.49$ are empirical coefficients. $q_s$ is calculated with respect to water phase or ice phase and environmental temperature.

We will deal with five types of cloud cover in this study: total cloud cover, which accounts for cloud cover over the entire atmospheric column and is most closely related to observational astronomy; layer cloud cover (low, mid, and high level), that divides according to the cloud-top pressure (680hPa, 440hPa, and $<440$hPa), which is particularly useful for observatories that mostly affected by a certain cloud type (e.g. high cloud for high altitude observatories); and convective cloud, which is mostly associated with convective weather that can be an unwarned threat for astronomical observation.

The GFS model divides the whole atmospheric column into 64 sub-layers for simulation, and cloud cover is derived under the assumption that clouds in all sub-layers for the corresponding layer are maximally randomly overlapped \citep{yan06}. The exception is convective cloud, which is derived based on the method proposed by \citet{pan95}.

\subsection{Observational Data}

Rather than using the ``traditional'' surface observation, we decide to use calibrated satellite observation in this study. The reason is that standard cloud observation is not practiced by a number of surface meteorological stations, as such observation needs to be carried out by expertise observers, therefore qualified observations are mostly limited to inhabited area with dense population, of which is avoid by most astronomical observatories. Satellite observation, on the other hand, has a better temporal and spatial coverage. Since it is carried out by robotic observers and is reduced following identical algorithms, it is also easier to determine the scale of uncertainty. A good source of such data is the International Satellite Cloud Climatology Project \citep[ISCCP; cf.][]{sch83,ros99}, which we will use for our study.

We retrieve the 3-hourly ISCCP D1 data from the ISCCP database for the same time period as the GFS data. The D1 dataset is in a spatial resolution of 280km and includes cloud cover data for total, low, mid and high cloud as well as convective cloud. The data is determined from raw satellite observation of cloud top pressure and optical thickness, and is in the same definition of the GFS data that we used. We then transform the ISCCP data from equal-area grid into equal-angle grid following the method by \citet{ros96}, to match the projection setup of the GFS data.

The uncertainty of ISCCP data is estimated to be $\sim0.15$ for individual cases and less than $\sim0.05$ for 30-day means \citep{sch83,ros99}. However, additional studies did reveal some observational tendencies for each cloud type:

\begin{enumerate}
  \item \citet{ros93} (surface observations): ISCCP is 0.10 too low over land (less in summer and more in winter);
  \item \citet{ros93,hah95} (surface observations): ISCCP misses some (up to 5\%) clouds at night;
  \item \citet{ros99} (Stratospheric Aerosol and Gas Experiments, High-Resolution Infrared Sounder, and surface observations): ISCCP high clouds are at least 0.05-0.10 too low;
  \item \citet{cur92,ros93} (surface and satellite observations): ISCCP total clouds for polar regions are 0.15-0.25 too low in summer and 0.05-0.10 too high in winter; and
  \item \citet{wie92} (satellite observations): overall tendencies of ISCCP low clouds are less than 0.1.
\end{enumerate}

As noted by \citet{cur92,ros93}, the ISCCP observations over the polar regions suffered from strong seasonal tendencies due to low visual and thermal contrast between surface and clouds; therefore, we will focus at the region between $60^{\circ}$S and $60^{\circ}$N in our study, despite we will still include results from polar regions in figures.

\section{Evaluation and Result}

\subsection{Evaluation Methodology}

We generate over 1 million GFS-ISCCP data pairs of every initialization, forecast time point and cloud type for the entire time period of interest. To get a more objective evaluation of the model forecast skill for total cloud cover, we compose three additional models that are to be compared with the ISCCP data: 

\begin{enumerate}
  \item Climatological model, which created by averaging ISCCP cloud data from July 2004 to June 2005. This model will be used to assess whether the GFS model is more skilled than statistical climatology forecast;
  \item Randomize model, which creates random series of pseudo global cloud fields under a uniform distribution. This model will be used to assess whether the GFS model is more skilled than unskilled guesses;
  \item Persistence model, which fix at the observation at $\tau=0h$ for a given initialization throughout the forecast period. This model will be used to compare the GFS model against a persistent ``guess''.
\end{enumerate}

Limiting by computational resource, we randomly choose July 2006 for such comparisons. We will show that 
monthly variation of forecast error is not significant in the period of study, so the July 2006 result is representative.

We use a different evaluation scheme for convective cloud. Although we are dealing with the term ``convective cloud cover'' or ``fractional convective cloud'', there is virtually no scientific/observational meaning in this term. It is due to the small scales of most convective clouds comparing with the spatial resolution of global model or satellite camera (which are mostly at tens of kilometers), so such cloud can only be represented in fractional numbers in model outputs or observations. Therefore, in addition to fractional comparisons, we also binary degenerate the modeling and observational data, so binary statistical indicators can be used to assess the forecast skill (see \S3.4.2).

\subsection{Total Cloud}

\subsubsection{General Forecast Accuracy}

\begin{figure}
\includegraphics[width=84mm]{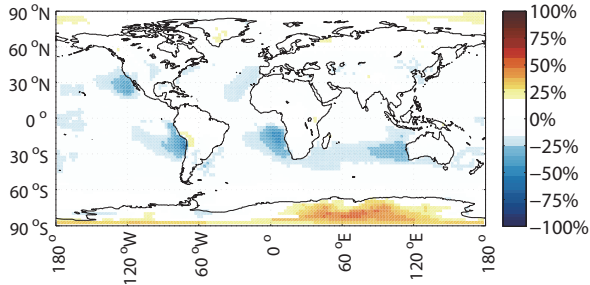}
\caption{The 3-year mean of forecast minus observation ($fc-obs$) distribution for total cloud cover forecast at $\tau=3h$ at 12Z initialization. Regions with $fc-obs$ beyond $\pm15\%$ are shaded in color.}
\label{fig1}
\end{figure}

\begin{figure}
\includegraphics[width=84mm]{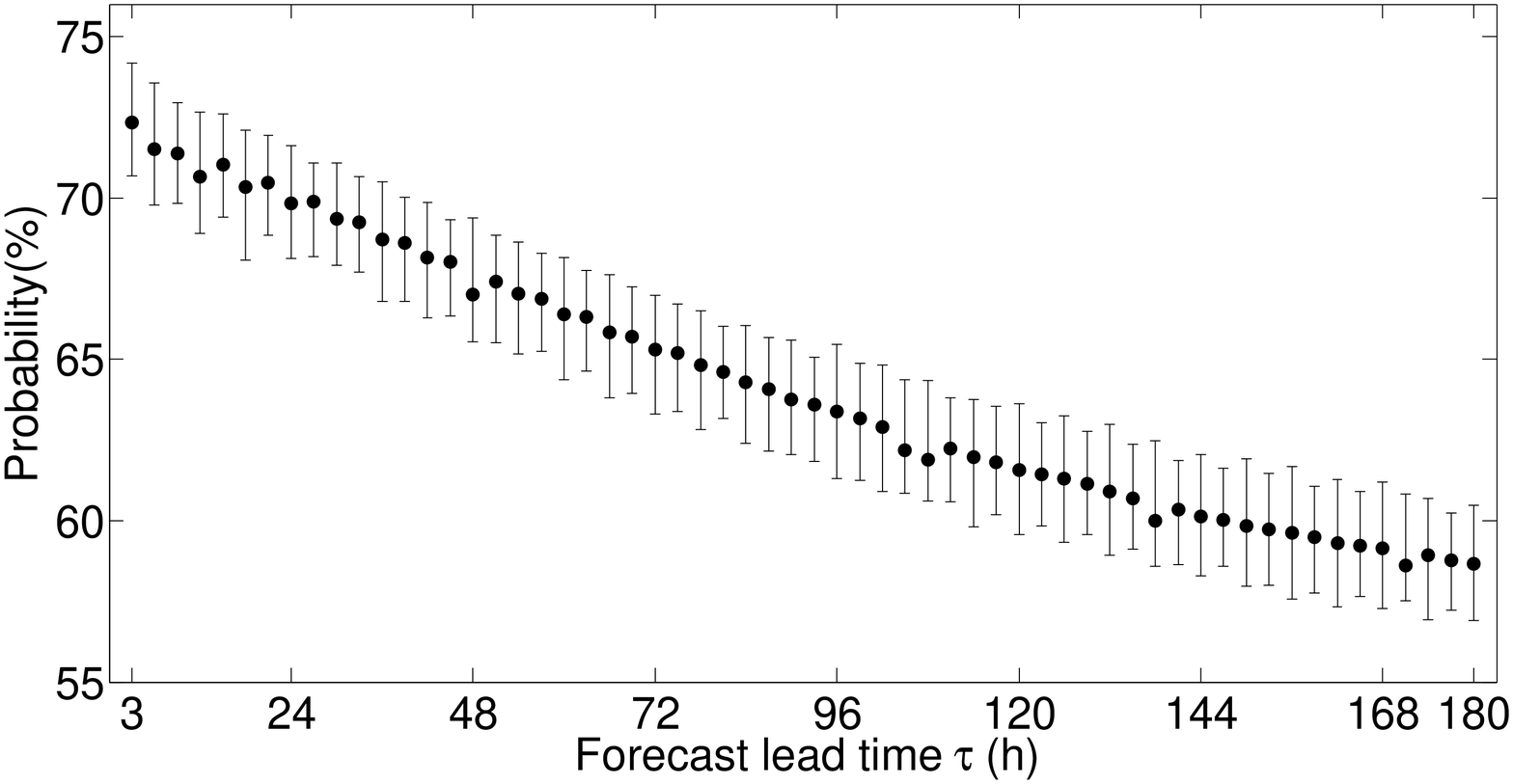}
\caption{The 3-year global mean (excluding polar regions) probability with $<30\%$ forecast error for $\tau=0-180h$ at 12Z initialization for total cloud cover forecast. Error bars represent standard variation.}
\label{fig2}
\end{figure}

Figure~\ref{fig1} shows the 3-year mean of forecast minus observation (abbreviated as $fc-obs$ below) of total cloud cover forecast at $\tau=3h$ at 12Z initialization; we notice that the $fc-obs$ setup for other time points are more or less the same and do not include them in the paper. Figure~\ref{fig2} shows the 3-year global mean (excluding polar regions) probability with $<30\%$ forecast error throughout the forecast period ($\tau$ from 0h to 180h) at 12Z initialization. Although root-mean-square error (RMSE) is commonly used on prediction evaluation, we notice that the preliminary RMSE figure behaves almost the same to the tendency figure (Figure~\ref{fig1}), which suggests that the main contribution of forecast RMSE to be persistent regional tendency rather than dispersion. Therefore, we argue that RMSE distribution and variation is representable with Figure~\ref{fig1} and Figure~\ref{fig2}.

In Table~\ref{tbl2} and Table~\ref{tbl3}, our analysis shows that the global mean $fc-obs$ (excluding polar regions) varies from -6.43\% to -8.93\% depending on initializations. Negative values are mostly contributed by low $fc-obs$ values over ocean, especially western coastal regions at mid latitude. The forecast over land matches the observation well (between $-0.5\%$ and $+1\%$ throughout the forecast period), but it may be positively biased according to the suggested underestimation of ISCCP data over land by \citet{ros93}. Meanwhile, the difference between each initialization is not significant comparing to the $fc-obs$ tendency (only $\sim3\%$ or less). We also find the negative tendency over the southern hemisphere is stronger than that of northern hemisphere.

\begin{table*}
 \centering
  \caption{Statistics of 3-year mean $fc-obs$ for total cloud cover forecast at 12Z initialization}
  \begin{tabular}{ccc}
  \hline
   Region & $\tau$ & Mean $fc$-$obs$ \\
 \hline
   Land & 3h & 0.28\% \\
        & 180h & 0.25\% \\
   Ocean & 3h & -10.31\% \\
         & 180h & -11.28\% \\
   Northern hemisphere & 3h & -5.85\% \\
                    & 180h & -6.34\% \\
   Southern hemisphere & 3h & -9.11\% \\
                    & 180h & -9.91\% \\
  \hline
\end{tabular}
\label{tbl2}
\end{table*}

\begin{table*}
 \centering
  \caption{Statistics of 3-year mean $fc-obs$ at each initialization}
  \begin{tabular}{cccc}
  \hline
   Initialization & $60^{\circ}$S to $60^{\circ}$N & $0^{\circ}$ to $60^{\circ}$N & $60^{\circ}$S to $0^{\circ}$ \\
 \hline
   00Z & -7.76\% & -5.42\% & -9.41\% \\
   06Z & -8.93\% & -7.92\% & -9.84\% \\
   12Z & -7.24\% & -5.33\% & -9.07\% \\
   18Z & -6.43\% & -4.78\% & -7.92\% \\
 \hline
\end{tabular}
\label{tbl3}
\end{table*}

Figure~\ref{fig2} shows the probability of $<30\%$ forecast error gradually decays from $\sim73\%$ at $\tau=3h$ to $\sim58\%$ at $\tau=180h$. One may argue that $\pm30\%$ has covered 3/5 of the possible range ($0\%-100\%$) which may lift the probability; however, cloud cover is not uniformly distributed: in a significant number of occasions, the cloud cover is either close to $0\%$ or $100\%$, so $\pm30\%$ is a fairly reasonable constraint. We will also show that the GFS model is solidly better than the random-guess model in later section.

\subsubsection{Daily and Seasonal Variation}

\begin{figure}
\includegraphics[width=84mm]{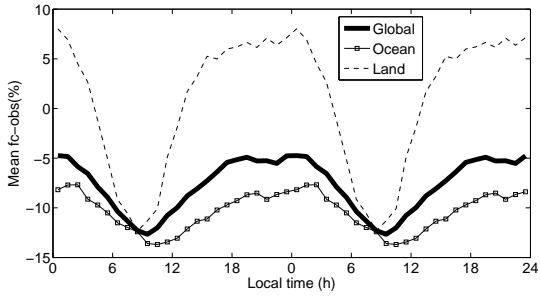}
\caption{3-year mean $fc-obs$ variation against local hour at 12Z initialization.}
\label{fig3}
\end{figure}

Since most astronomical optical observations are conducted in night hours, we are interested in examining the daily variation of forecast accuracy. We divide the entire globe into 24 time zones with equally longitudinal spacing, and average the $fc-obs$ series with respect to local hour in each time zone. As illustrated in Figure~\ref{fig3}, the daily $fc-obs$ value varies between $-15\%$ to $-5\%$ for the entire globe (excluding polar regions) and ocean, but for land it varies from $-13\%$ in the morning to $+5\%$ to $+10\%$ in night hours. However, considering a $\sim7\%$ underestimation of ISCCP data over land in day and $\sim12\%$ underestimation in night \citep{ros93}, the actual $fc-obs$ could be as low as $-25\%$ in day but near $0\%$ in night.

\begin{figure}
\includegraphics[width=84mm]{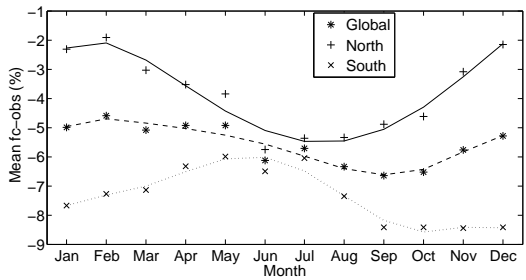}
\caption{3-year mean monthly $fc-obs$ variation at 12Z initialization.}
\label{fig4}
\end{figure}

\begin{figure}
\includegraphics[width=84mm]{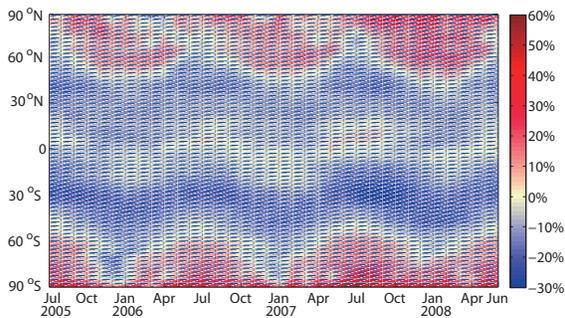}
\caption{Monthly zonal mean $fc-obs$ variation throughout the period of study at 12Z initialization.}
\label{fig5}
\end{figure}

Figure~\ref{fig4} shows a clear seasonal variation of mean $fc-obs$ that reaches maximum in winter and minimum in summer in respective hemisphere, but it might not reflect the actual situation, as the study of \citet{ros93} has indicated that seasonal difference of ISCCP tendency can be as large as 9\%, with a reverse maximum-minimum setup ($-6\%$ in summer and $-15\%$ in winter) than Figure~\ref{fig4}. In such a case, it is possible that the GFS model forecast is closer to actual situation in summer, rather than in winter as suggested in the figure. We also attempt to identify any annual variation (Figure~\ref{fig5}), but no significant feature can be noted, possibly due to insufficient temporal coverage.

\subsubsection{Comparison with Climatology, Randomize and Persistence Models}

\begin{figure}
\includegraphics[width=84mm]{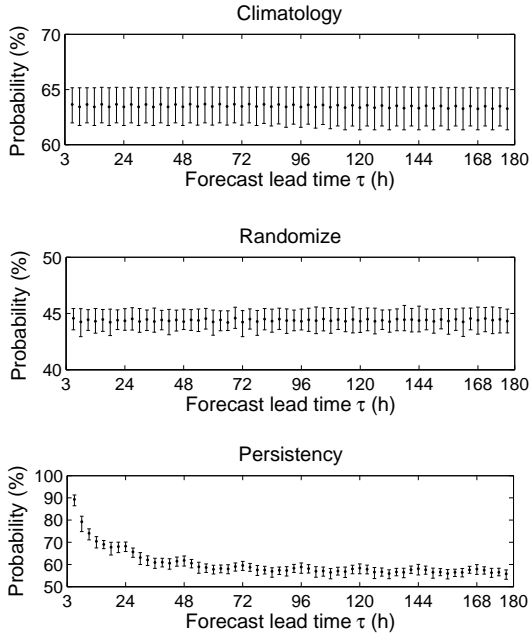}
\caption{The global mean (excluding polar regions) probability variation with $<30\%$ forecast error for climatology, randomize and persistence models for July 2006. Error bars represent standard deviation.}
\label{fig6}
\end{figure}

As described before, we generated three additional models and will use them to compare with the ISCCP data for July 2006. The result is shown in Figure~\ref{fig6} and, in our opinion, is comparable with Figure~\ref{fig2} despite different temporal coverage, as we have shown the seasonal variation to be insignificant comparing to overall $fc-obs$ tendency.

We can see the GFS model is superior than all three other models in most time points. Statistically, the persistence model is best of all for $\tau<6h$, but this is not meaningful as the GFS model data is not available after approximately 4-5h of the respective initialization time due to computational layover. We note that the ISCCP climatology model becomes better than the GFS model after $\tau\sim120h$, and that both models are more skilled than random guess at all times. From this result, we can conclude that the GFS model is skilled, and performs better than the ISCCP climatology model until $\tau\sim120h$.

\subsection{Layer Cloud}

\begin{figure}
\includegraphics[width=84mm]{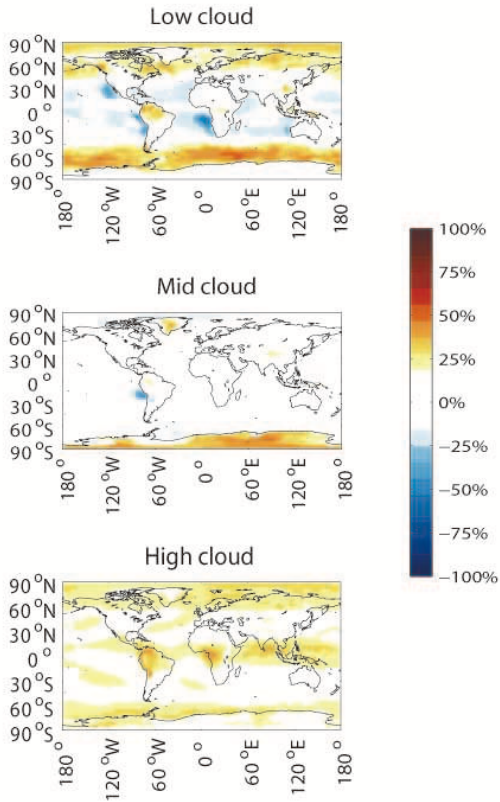}
\caption{The 3-year mean $fc-obs$ distribution for low, mid and high cloud cover forecast at $\tau=3h$ at 12Z initialization. Regions with $fc$-$obs$ beyond $\pm15\%$ are shaded in color.}
\label{fig7}
\end{figure}

\begin{figure}
\includegraphics[width=84mm]{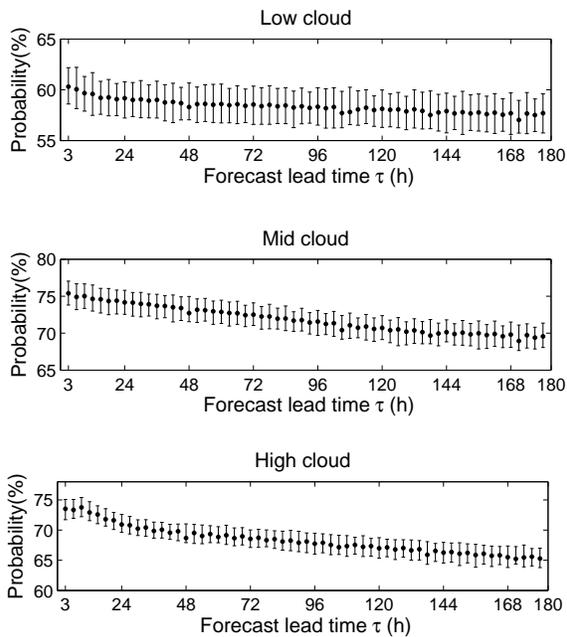}
\caption{The 3-year global mean (excluding polar regions) probability variation with $<30\%$ forecast error for $\tau=0-180h$ for low, mid and high cloud cover forecast at 12Z initialization. Error bars represent standard deviation.}
\label{fig8}
\end{figure}

\begin{figure}
\includegraphics[width=84mm]{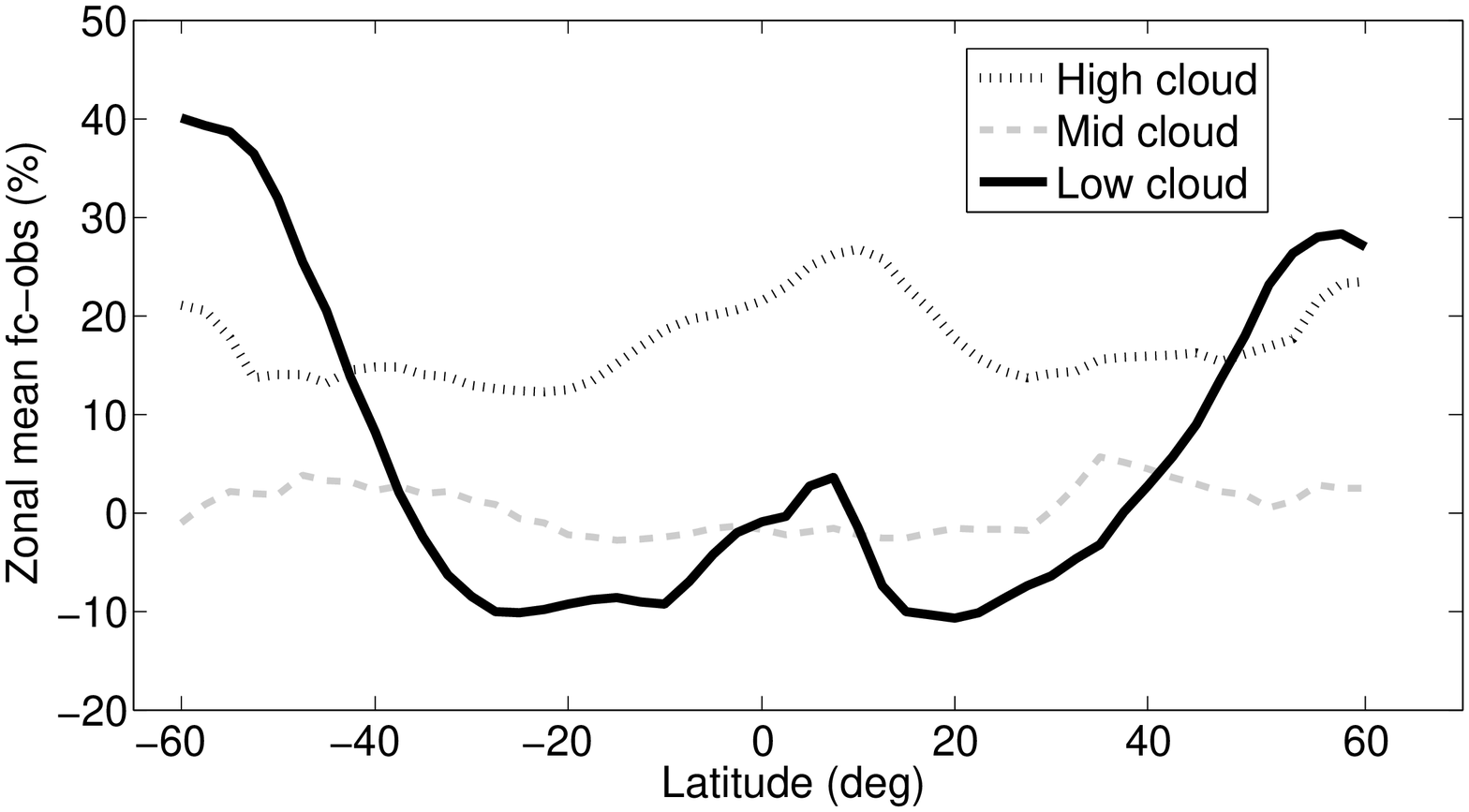}
\caption{The 3-year mean zonal $fc-obs$ for low, mid and high cloud at 12Z initialization.}
\label{fig9}
\end{figure}

The evaluation result (Table~\ref{tbl4}, Figure~\ref{fig7}, Figure~\ref{fig8} and Figure~\ref{fig9}) has revealed the global overestimation of high cloud from the GFS model, but this result is affected by underestimation of similar scale in ISCCP data \citep{ros99} and is questionable. We can also identify underestimation of low cloud off the west coast of major continents at mid latitude, which seems to be the major contribution to the underestimation of total cloud in these regions. At high latitude, the model tends to overestimate the low cloud.

\begin{table*}
 \centering
  \caption{Statistics of 3-year mean $fc-obs$ for low, mid and high cloud at 12Z initialization}
  \begin{tabular}{ccc}
  \hline
   Cloud type & Region & 3-year mean $fc-obs$ \\
 \hline
   Low cloud & $60^{\circ}$S to $60^{\circ}$N & +3.08\% \\
    & $0^{\circ}$ to $60^{\circ}$N & +1.60\% \\
    & $60^{\circ}$S to $0^{\circ}$ & +4.35\% \\
    & Land & +7.05\% \\
    & Ocean & +1.82\% \\
   Mid cloud & $60^{\circ}$S to $60^{\circ}$N & -1.21\% \\
    & $0^{\circ}$ to $60^{\circ}$N & -0.59\% \\
    & $60^{\circ}$S to $0^{\circ}$ & -1.80\% \\
    & Land & +1.63\% \\
    & Ocean & +1.60\% \\
   High cloud & $60^{\circ}$S to $60^{\circ}$N & +17.11\% \\
    & $0^{\circ}$ to $60^{\circ}$N & +18.79\% \\
    & $60^{\circ}$S to $0^{\circ}$ & +15.67\% \\
    & Land & +15.92\% \\
    & Ocean & +16.85\% \\
  \hline
\end{tabular}
\label{tbl4}
\end{table*}

The probability of $<30\%$ error forecast for low cloud is significantly lower than other two clouds (Figure~\ref{fig8}), but does not appear to affect the total cloud at a small $\tau$. This behavior, together with the unexpected negative $fc-obs$ for total cloud (while $fc-obs$ for layer clouds are mostly positive), could indicate that the assumption of maximally randomly overlaps (see \S2.1) between each cloud layers may not apply at all times.

\subsection{Convective Cloud}

\subsubsection{Fractional Evaluation}

\begin{figure}
\includegraphics[width=84mm]{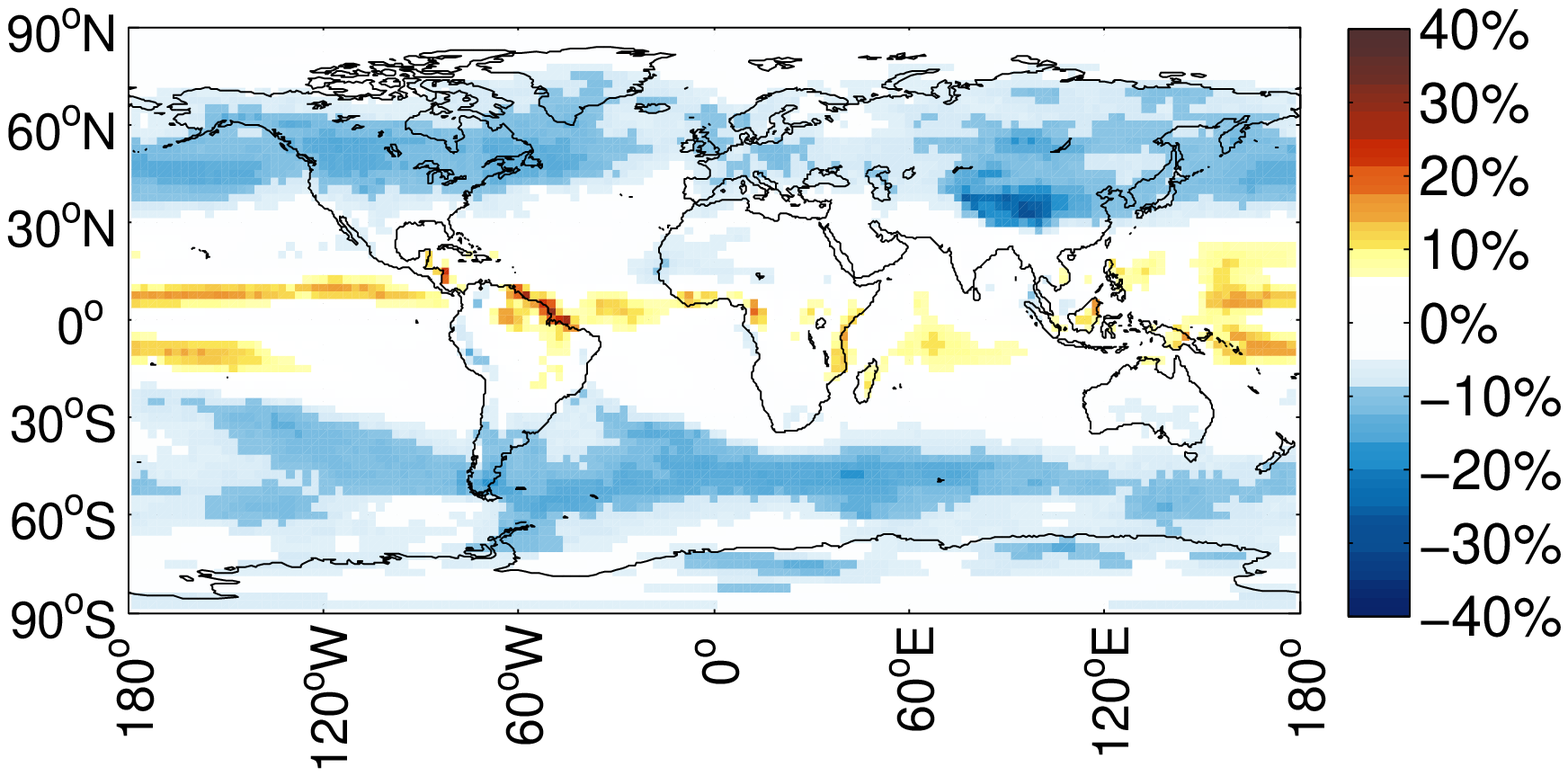}
\caption{The 3-year mean $fc-obs$ for convective cloud forecast at $\tau=3h$ of 12Z initialization. Regions with $fc-obs$ beyond $\pm15\%$ are shaded in color.}
\label{fig10}
\end{figure}

Direct fractional comparison (Table~\ref{tbl5} and Figure~\ref{fig10}) shows that model overestimation only occurs in tropical area (within $15^{\circ}$ latitude in both hemispheres), while moderate to strong underestimation (mostly $10-20\%$) occurs in mid-to-high latitude area. Considering the fact that convective weather is frequent in tropic area throughout the year but is relatively rare in mid-to-high latitude area, we may conclude that the current model misses and/or underestimates a significant fraction of convective cloud in mid-to-high latitude area, while it tends to overestimate the number and/or intensity of convective cloud in tropical area.

\begin{table*}
 \centering
  \caption{Statistics of 3-year mean $fc-obs$ for convective cloud at 12Z initialization}
  \begin{tabular}{ccc}
  \hline
   Region & $\tau$ & 3-year mean $fc-obs$ \\
 \hline
   Land & 3h & -5.11\% \\
    & 180h & -5.08\% \\
   Ocean & 3h & -3.39\% \\
    & 180h & -4.37\% \\
   Northern hemisphere & 3h & -3.67\% \\
    & 180h & -4.41\% \\
   Southern hemisphere & 3h & -3.65\% \\
    & 180h & -4.88\% \\
 \hline
\end{tabular}
\label{tbl5}
\end{table*}

\subsubsection{Binary Evaluation}

\begin{figure}
\includegraphics[width=84mm]{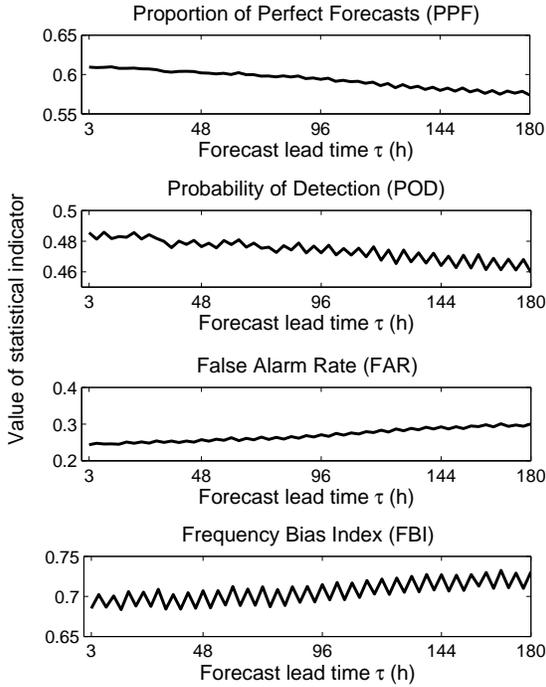}
\caption{The global mean variations (polar regions excluded) of four statistical indicators for convective cloud forecast at 12Z initialization throughout the forecast period.}
\label{fig11}
\end{figure}

An alternative way to evaluate the GFS convective cloud forecast is to degenerate the forecast and observation into binary value, so we can focus on the occurrence of such cloud rather than a mixture of occurrence and intensity. As discussed in earlier section, we set a cut-off limit at 1\% for convective cloud fraction to put both the modeling and observational data into the categories of ``convective cloud'' and ``no convective cloud''. We use the following statistical indicators for evaluation: Proportion of Perfect Forecasts (PPF), Probability of Detection (POD), False Alarm Rate (FAR) and Frequency Bias Index (FBI). In the following expressions, $H$ indicates ``hits'' (forecasted and observed), $F$ indicates ``false alarms'' (forecasted but not observed), $M$ indicates ``missed'' (not forecasted but observed), and $Z$ indicates to not-forecasted and not-observed events. The result is shown in Figure~\ref{fig11}.

\begin{displaymath}
PPF=\frac{H+Z}{H+F+M+Z}
\end{displaymath}
\begin{displaymath}
POD=\frac{H}{H+M}
\end{displaymath}
\begin{displaymath}
FAR=\frac{F}{F+Z}
\end{displaymath}
\begin{displaymath}
FBI=\frac{H+F}{H+M}
\end{displaymath}

As we have shown that tropical overestimation only composed a relatively small fraction in the sample, the setup of Figure~\ref{fig11} more or less represents the situation of mid-to-high latitude area. This is affirmed by the FBI value which lies way below 1, suggesting a global underestimation of convective cloud.

The PPF varies around 0.6 and creates a decent picture of the forecast ability, but the rare occurrence of convective cloud in most areas will lead to a significant fraction of PPF being contributed from ``Z'' events (not-forecasted and not-observed events), so we must take POD and FAR into account for a unbiased view. From POD we notice that only slightly less than half of the convective cloud can be detected by the model. Since convective cloud is most commonly seen in tropical area, the POD for mid-to-high latitude area must be lower. However, globally speaking, the model is still well skilled as the FAR is about 2 times lower than the POD.

\section{Concluding Remarks}

To study the reliability of cloud forecast from the GFS model as a reference of astronomical observations, we analyzed a 3-year sample composed with the GFS modeling data and satellite observation data from ISCCP. The results are summarized as follow:

\begin{enumerate}
  \item For total cloud cover forecast, there is a slight global underestimation from the GFS model, but is more or less within the observational uncertainty of the ISCCP data. For local night hours, the model forecast roughly agrees with the observation if taking account the observational tendency of ISCCP data as suggested by earlier studies. The global mean probability (excluding polar regions) of $<30\%$ forecast error gradually declines from 73\% to 58\% from $\tau=3h$ to $\tau=180h$. Further investigation suggests that climatology model based on ISCCP observation overtakes the GFS model after $\tau\sim120h$, but both models are significantly more skilled than random guesses.
  \item We found a strong underestimation of low clouds in subtropical regions off the west coast of both hemispheres. Overestimation of low clouds in high latitude areas can also be identified. We found some $15\%$ globally overestimation of high clouds, but is most likely compromised by a similar scale of observational tendency in the ISCCP data. We noted the inconsistency of mean global forecast errors between total and layer clouds, which might be due to the layer overlapping assumption built in the model.
  \item For convective cloud forecast, the GFS model tends to overestimate the occurrence and/or intensity of convective cloud in tropical areas but tends to underestimate that in subtropical and high latitude areas. We found the GFS model can only identify less than half of the convective cloud globally. However, the convective cloud forecast is still skilled, as the detection rate is about two times higher than the false alarm rate, leading to a proportion of prefect forecast of $\sim0.6$.
\end{enumerate}

In all, we observed a good overall consistency between the GFS model forecast and ISCCP observation throughout the time period of interest. For total cloud cover, our result suggested a satisfactory performance of the GFS model for the need of observation scheduling up to a week ahead. However, for layer and convective cloud, which can be considerably important for observatories located in certain environment (for example, high cloud for high altitude observatories), the success rate of the GFS model is relatively less satisfying than that of total cloud forecast.

\section*{Acknowledgment}

We would like to thank several anonymous reviewers as well as Fangling Yang and William Rossow for their constructive comments and discussions that help us to make significant improvements of this work. We also would like to thank Chen Junwen from Atmospheric Exploration Laboratory (EESAEL) at Sun Yat-sen University, Cui Chenzhou (National Astronomical Observatory), Lin Qing and Tang Haiming (Shanghai Astronomical Observatory), for providing assistance on computational resources. We would like to especially thank the ``meteo cat'' who had visited the EESAEL a few times and brought some joys to us. The data analysis of this work was accomplished with the computer resource at EESAEL. The GFS data were obtained from NOMADS (National Operational Model Archive and Distribution System) at NOAA/NESDIS/NCDC. The ISCCP D1 data were obtained from the International Satellite Cloud Climatology Project data archives at NOAA/NESDIS/NCDC Satellite Services Group, ncdc.satorder@noaa.gov, on July to October, 2010.

\bibliographystyle{mn2e}
\bibliography{man}

\label{lastpage}

\end{document}